\documentclass[aps,prb,reprint,superscriptaddress]{revtex4-2}
\usepackage[T1]{fontenc}
\usepackage{times}
\usepackage{graphicx,color}
\usepackage{amsfonts,amsmath,amssymb,amsbsy}
\usepackage[colorlinks=true, linkcolor=blue, citecolor=blue, urlcolor=blue]{hyperref}
\let\mathbf=\boldsymbol
\def\blue#1{\textcolor{blue}{#1}}
\def\emph#1{\textcolor{magenta}{#1}}
\begin{document}
%%%%%%%%%%%%%%%%%%%%%%%%%%%%%%%%%%%%%%%%%%%%%%%%%%%%%%%%%%%%

\title{{\Large Bifurcation of a Topological Skyrmion String}}

\author{Jing Xia}
%\email[Email:~]{jingxia0817@gmail.com}
\thanks{These authors contributed equally to this work.}
\affiliation{College of Physics and Electronic Engineering, Sichuan Normal University, Chengdu 610068, China}

\author{Xichao Zhang}
%\email[Email:~]{zhangxichao_jsps@shinshu-u.ac.jp}
\thanks{These authors contributed equally to this work.}
\affiliation{Department of Electrical and Computer Engineering, Shinshu University, 4-17-1 Wakasato, Nagano 380-8553, Japan}

\author{Oleg A. Tretiakov}
%\email[Email:~]{o.tretiakov@unsw.edu.au}
\affiliation{School of Physics, The University of New South Wales, Sydney 2052, Australia}

\author{Hung T. Diep}
%\email[Email:~]{diep@u-cergy.fr}
\affiliation{Laboratoire de Physique Th{\'e}orique et Mod{\'e}lisation, CY Cergy Paris Universit{\'e}, 95302 Cergy-Pontoise Cedex, France}

\author{Jinbo Yang}
%\email[Email:~]{jbyang@pku.edu.cn}
\affiliation{State Key Laboratory for Mesoscopic Physics, School of Physics, Peking University, Beijing, 100871, China}

\author{\\ Guoping Zhao}
%\email[Email:~]{zhaogp@uestc.edu.cn}
\affiliation{College of Physics and Electronic Engineering, Sichuan Normal University, Chengdu 610068, China}

\author{Motohiko Ezawa}
\email[]{ezawa@ap.t.u-tokyo.ac.jp}
\affiliation{Department of Applied Physics, The University of Tokyo, 7-3-1 Hongo, Tokyo 113-8656, Japan}

\author{Yan Zhou}
\email[]{zhouyan@cuhk.edu.cn}
\affiliation{School of Science and Engineering, The Chinese University of Hong Kong, Shenzhen, Guangdong 518172, China}

\author{Xiaoxi Liu}
\email[]{liu@cs.shinshu-u.ac.jp}
\affiliation{Department of Electrical and Computer Engineering, Shinshu University, 4-17-1 Wakasato, Nagano 380-8553, Japan}

%-%-%-%-%-%-%-%-%-%-%-%-%-%-%-%-%-%-%-%-%-%-%-%-%-%-%-%-%-%-%
\begin{abstract}
Manipulation of three-dimensional (3D) topological objects is of both fundamental interest and practical importance in many branches of physics. Here, we show by spin dynamics simulations that the bifurcation of a 3D skyrmion string in a layered frustrated system could be induced by the dampinglike spin-orbit torque. The bifurcation of a skyrmion string happens when the skyrmion string carries a minimal topological charge of $Q=2$. We demonstrate that three types of bifurcations could be realized by applying different current injection geometries, which lead to the transformation from I-shaped skyrmion strings to Y-, X-, and O-shaped ones. Besides, different branches of a bifurcated skyrmion string may merge into an isolated skyrmion string spontaneously. The mechanism of bifurcation should be universal to any skyrmion strings with $Q\geq 2$ in the layered frustrated system and could offer a general approach to manipulate 3D stringlike topological objects for spintronic functions.
\end{abstract}
%-%-%-%-%-%-%-%-%-%-%-%-%-%-%-%-%-%-%-%-%-%-%-%-%-%-%-%-%-%-%

%\date{\today}
\date{May 19, 2022}

%\preprint{\textsl{Revised Version - Transferred to PRB}}

%\keywords{skyrmion, skyrmionium, bimeron, bimeronium, frustrated magnet, spintronics, micromagnetics}
%\pacs{75.10.Hk, 75.10.Jm, 75.70.Ak, 75.70.Kw, 75.78.-n, 12.39.Dc}
% Classical spin models:                                     75.10.Hk
% Quantized spin models, including quantum spin frustration: 75.10.Jm
% Magnetic properties of monolayers and thin films:          75.70.Ak
% Domain structure (magnetic bubbles and vortices):          75.70.Kw
% Magnetization dynamics:                                    75.78.-n
% Skyrmions:                                                 12.39.Dc

\maketitle

%-%-%-%-%-%-%-%-%-%-%-%-%-%-%-%-%-%-%-%-%-%-%-%-%-%-%-%-%-%-%
%\section{Introduction}
%\label{se:Introduction}
%-%-%-%-%-%-%-%-%-%-%-%-%-%-%-%-%-%-%-%-%-%-%-%-%-%-%-%-%-%-%

\textit{Introduction.}
The interdisciplinary interaction between physics and topology has created many hot topics in recent years, including the topological magnetism~\cite{Nagaosa_NNANO2013,Mochizuki_Review,Wiesendanger_Review2016,Finocchio_JPD2016,Kang_PIEEE2016,Kanazawa_AM2017,Wanjun_PHYSREP2017,Fert_NATREVMAT2017,Everschor_JAP2018,Zhang_JPCM2020,Fujishiro_2020,Gobel_PP2021,Reichhardt_2021,Back_JPD2020} and topological photonics~\cite{Ozawa_2019,Shen_2021}.
Particularly, the manipulation of three-dimensional (3D) particlelike topological objects in physical world is of special fundamental interest and technological importance~\cite{Everschor_JAP2018,Zhang_JPCM2020,Gobel_PP2021,Fujishiro_2020,Reichhardt_2021,Shen_2021,Back_JPD2020}.
Representative topologically nontrivial quasiparticles are magnetic spin structures carrying nonzero topological charges~\cite{Nagaosa_NNANO2013,Mochizuki_Review,Wiesendanger_Review2016,Finocchio_JPD2016,Kang_PIEEE2016,Kanazawa_AM2017,Wanjun_PHYSREP2017,Fert_NATREVMAT2017,Everschor_JAP2018,Zhang_JPCM2020,Fujishiro_2020,Gobel_PP2021,Reichhardt_2021,Back_JPD2020}, such as the magnetic skyrmion~\cite{Roszler_NATURE2006,Bogdanov_1989}.
Topological spin structures can be classified into one-dimensional, two-dimensional (2D) and 3D categories~\cite{Everschor_JAP2018,Zhang_JPCM2020,Gobel_PP2021,Fujishiro_2020,Reichhardt_2021,Back_JPD2020}, while some 3D topological spin structures can be formed by 2D ones in exchange-coupled bilayer and multilayer systems~\cite{Wanjun_SCIENCE2015,Woo_NM2016,ML_NN2016,Boulle_NN2016,Soumyanarayanan_NM2017,Mandru_NC2020,Zeissler_NC2020,Litzius_NPHYS2017}.
Pure 2D and quasi-2D topological spin structures are usually simple and rigid~\cite{Lin_PRB2013,Reichhardt_PRL2015,Reichhardt_PRB2015}, of which the dynamics can be well described and controlled, making them promising candidates for spintronic applications~\cite{Nagaosa_NNANO2013,Mochizuki_Review,Wiesendanger_Review2016,Finocchio_JPD2016,Kang_PIEEE2016,Kanazawa_AM2017,Wanjun_PHYSREP2017,Fert_NATREVMAT2017,Everschor_JAP2018,Zhang_JPCM2020,Fujishiro_2020,Gobel_PP2021,Reichhardt_2021,Back_JPD2020}.

However, recent studies suggest that 3D topological spin structures, such as skyrmion strings~\cite{Sutcliffe_2017,Kagawa_2017,Yokouchi_2018,Sohn_2019,Koshibae_2019,Seki_2020,Koshibae_2020,Yu_2020,Birch_NC2021,Kravchuk_2020,Xing_2020,Seki_2021,Zhang_2021,Tang_2021,Xia_2021,Zheng_2021,Marrows_2021} and hopfions~\cite{Sutcliffe_2018,Wang_PRL2019,Liu_PRL2020,Voinescu_PRL2020,Kent_NC2021}, also have great potential to be used as essential components in spintronic applications.
For example, a skyrmion string can be utilized as a transport channel for magnons~\cite{Seki_2020,Kravchuk_2020,Xing_2020}.
A hopfion can carry information in a nonvolatile manner and can be displaced by spin currents with no undesirable Hall effects~\cite{Wang_PRL2019}.
In addition, 3D topological spin structures usually have more degrees of freedom that can be manipulated by external stimuli, which could generate complicated but fascinating physical phenomena that may have implications for applications.

Most 3D topological spin structures exist in bulk and layered magnetic systems with chiral or frustrated exchange interactions~\cite{Sutcliffe_2017,Kagawa_2017,Yokouchi_2018,Sohn_2019,Koshibae_2019,Seki_2020,Koshibae_2020,Yu_2020,Birch_NC2021,Kravchuk_2020,Xing_2020,Seki_2021,Zhang_2021,Tang_2021,Xia_2021,Zheng_2021,Sutcliffe_2018,Wang_PRL2019,Liu_PRL2020,Voinescu_PRL2020,Kent_NC2021,Zhang_JPCM2020,Gobel_PP2021,Fujishiro_2020,Back_JPD2020}.
The frustrated magnetic systems with competing exchange interactions~\cite{Diep_Entropy2019,Batista_2016} can host both 2D skyrmions~\cite{Leonov_NCOMMS2015,Lin_PRB2016A,Rozsa_PRL2016,Leonov_NCOMMS2017,Xichao_NCOMMS2017,Yuan_PRB2017,Ritzmann_NE2018,Kurumaji_SCIENCE2019,Xia_PRApplied2019,Gobel_PRB2019} and 3D skyrmion strings~\cite{Lin_PRL2018,Sutcliffe_2017,Zhang_2021}, which show unique dynamical properties that cannot be found in other systems.
For example, a 2D frustrated skyrmion with a topological charge of $Q=1$ driven by the dampinglike spin-orbit torque, where $Q=-{\frac{1}{4\pi}}\int\boldsymbol{m}(\boldsymbol{r})\cdot\left[\partial_{x}\boldsymbol{m}(\boldsymbol{r})\times\partial_{y}\boldsymbol{m}(\boldsymbol{r})\right]d^{2}\boldsymbol{r}$ with $\boldsymbol{m}$ being the normalized spin, could move along a circular path accompanied by the rotation of its helicity~\cite{Lin_PRB2016A,Xichao_NCOMMS2017}.
It is envisioned that the manipulation of a 3D frustrated skyrmion string may also uncover unexpected physical properties of 3D topological spin structures.
In this Letter, we computationally demonstrate the current-induced bifurcation of a 3D skyrmion string.
The physical bifurcation of a skyrmion string could result in the formation of a complex skyrmion string with multiple topological branches, which is a novel phenomenon that cannot be found in 2D or quasi-2D systems.

%%%%%%%%%%%%%%%%%%%%%%%%%%%%%%%%%%%%%%%%%%%%%%%%%%%%%%%%%%%%
\begin{figure*}[t]
\centerline{\includegraphics[width=1.00\textwidth]{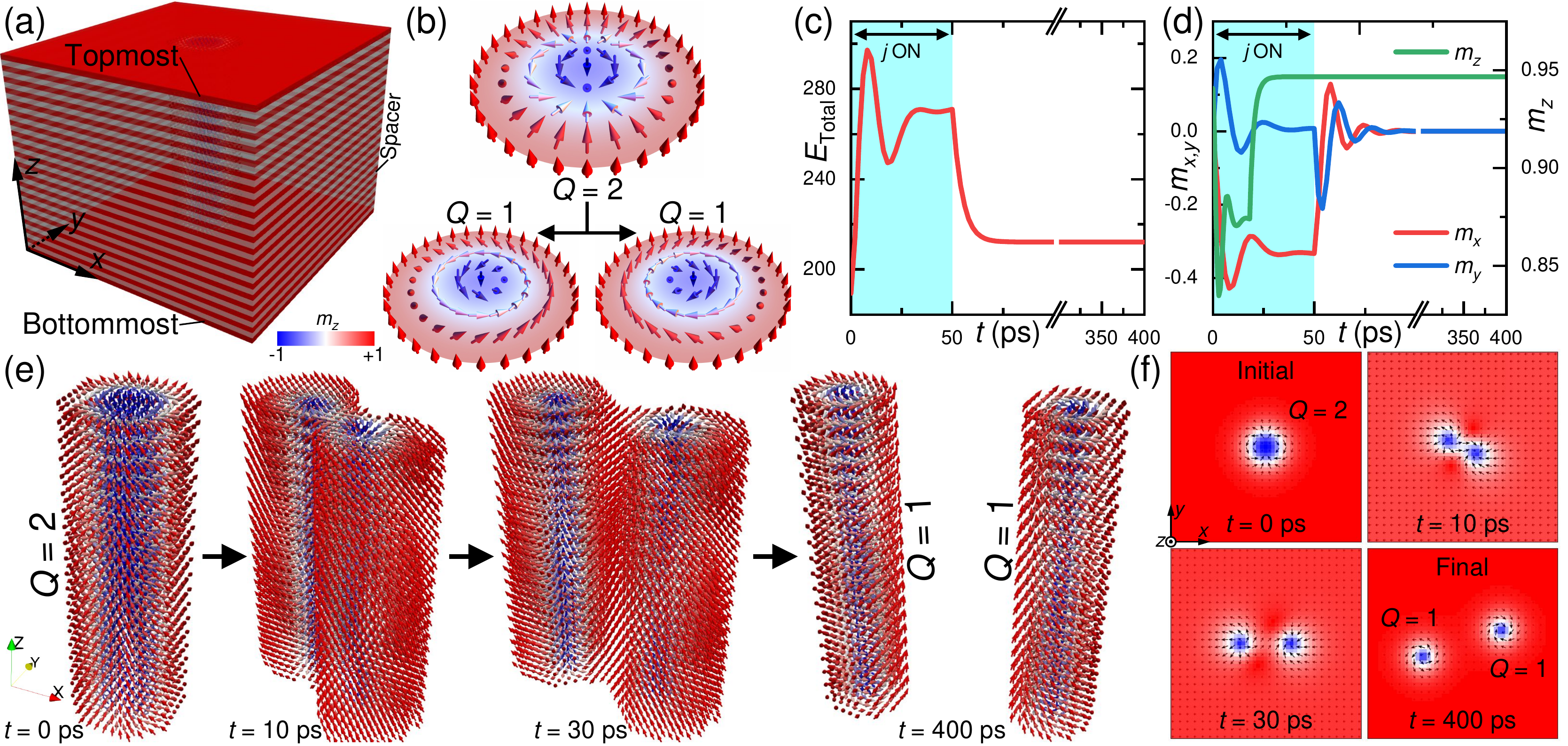}}
\caption{%
Separation of an I-shaped skyrmion string with $Q=2$.
(a) Schematic of the model, where $20$ frustrated FM layers are exchange-coupled through FM interlayer exchange interactions. FM layers and nonmagnetic spacers are indicated by red and gray colors, respectively.
(b) Illustrations of a 2D skyrmion with $Q=2$ and two 2D skyrmions with $Q=1$ and opposite helicities.
(c) Total energy $E_{\text{Total}}$ as a function of time for the separation of an I-shaped skyrmion string with $Q=2$ into two I-shaped skyrmion strings with $Q=1$. The energy is given in units of $J_{1}=1$. The driving current is turned on for $t=0-50$ ps, indicated by the cyan background.
(d) Time-dependent in-plane spin components $m_{x,y}$ and out-of-plane spin component $m_z$ of the system for the separation of the skyrmion string.
(e) Illustrations showing the separation and relaxation of the skyrmion string at selected times. Only the areas with $m_z\leq 0.8$ are visible. Each arrow stands for a spin.
(f) Top-view snapshots of the topmost FM layer at selected times corresponding to (e). Black arrows represent spins with a subsample rate of $2$.
}
\label{FIG1}
\end{figure*}
%%%%%%%%%%%%%%%%%%%%%%%%%%%%%%%%%%%%%%%%%%%%%%%%%%%%%%%%%%%%

%-%-%-%-%-%-%-%-%-%-%-%-%-%-%-%-%-%-%-%-%-%-%-%-%-%-%-%-%-%-%
%\section{Methods}
%\label{se:Methods}
%-%-%-%-%-%-%-%-%-%-%-%-%-%-%-%-%-%-%-%-%-%-%-%-%-%-%-%-%-%-%

\textit{Model.}
We consider $20$ weakly exchange-coupled ferromagnetic (FM) layers with exchange frustration [Fig.~\ref{FIG1}(a)].
Each FM layer is described by a $J_{1}$-$J_{2}$-$J_{3}$ classical Heisenberg model on a simple square lattice~\cite{Batista_2016,Lin_PRB2016A,Xichao_NCOMMS2017}, where $J_1$, $J_2$, and $J_3$ denote the FM nearest-neighbor (NN), antiferromagnetic (AFM) next-NN (NNN), and AFM next-NNN intralayer exchange interactions, respectively.
The total Hamiltonian is given in Supplemental Note \blue{1}~\cite{SI}.
We note that two NN FM layers in our model are separated by either a nonmagnetic insulating or a hybrid metal-insulator spacer layer to ensure the weak FM interlayer coupling $J_{\text{inter}}$~\cite{RKKY1,RKKY2,RKKY3} and the spatially inhomogeneous current injection; see Supplemental Note \blue{2}~\cite{SI}.
The spin dynamics is described by the Landau-Lifshitz-Gilbert equation augmented with the dampinglike spin-orbit torque $\boldsymbol{\tau}_{\text{d}}=\frac{u}{b}\left(\boldsymbol{m}\times\boldsymbol{p}\times\boldsymbol{m}\right)$~\cite{OOMMF}, which could be generated via the spin Hall effect~\cite{Sinova_SHE,Hoffmann_SHE},
\begin{equation}
\frac{d\boldsymbol{m}}{dt}=-\gamma_{0}\boldsymbol{m}\times\boldsymbol{h}_{\rm{eff}}+\alpha\left(\boldsymbol{m}\times\frac{d\boldsymbol{m}}{dt}\right)+\boldsymbol{\tau}_{\text{d}},
\label{eq:LLG}
\end{equation}
where $\boldsymbol{h}_{\rm{eff}}=-\frac{1}{\mu_{0}M_{\text{S}}}\cdot\frac{\delta\mathcal{H}}{\delta\boldsymbol{m}}$ is the effective field,
$u=\left|\left(\gamma_{0}\hbar/\mu_{0}e\right)\right|\cdot\left(j\theta_{\text{SH}}/2 M_{\text{S}}\right)$ is the spin-orbit torque coefficient,
$\mu_{0}$ is the vacuum permeability constant,
$M_{\text{S}}$ is the saturation magnetization,
$t$ is the time,
$\alpha$ is the Gilbert damping parameter,
$\gamma_0$ is the absolute gyromagnetic ratio,
$\hbar$ is the reduced Planck constant, $e$ is the electron charge, $b$ is the single FM layer thickness, $j$ is the current density, and $\theta_{\text{SH}}$ is the spin Hall angle.
The spin polarization orientation $\boldsymbol{p}=+\hat{y}$.
The lattice constant is $a=0.4$ nm, and the mesh size is $a^3$.
The default parameters are given in Supplemental Note \blue{1}~\cite{SI}.

%%%%%%%%%%%%%%%%%%%%%%%%%%%%%%%%%%%%%%%%%%%%%%%%%%%%%%%%%%%%
\begin{figure*}[t]
\centerline{\includegraphics[width=1.00\textwidth]{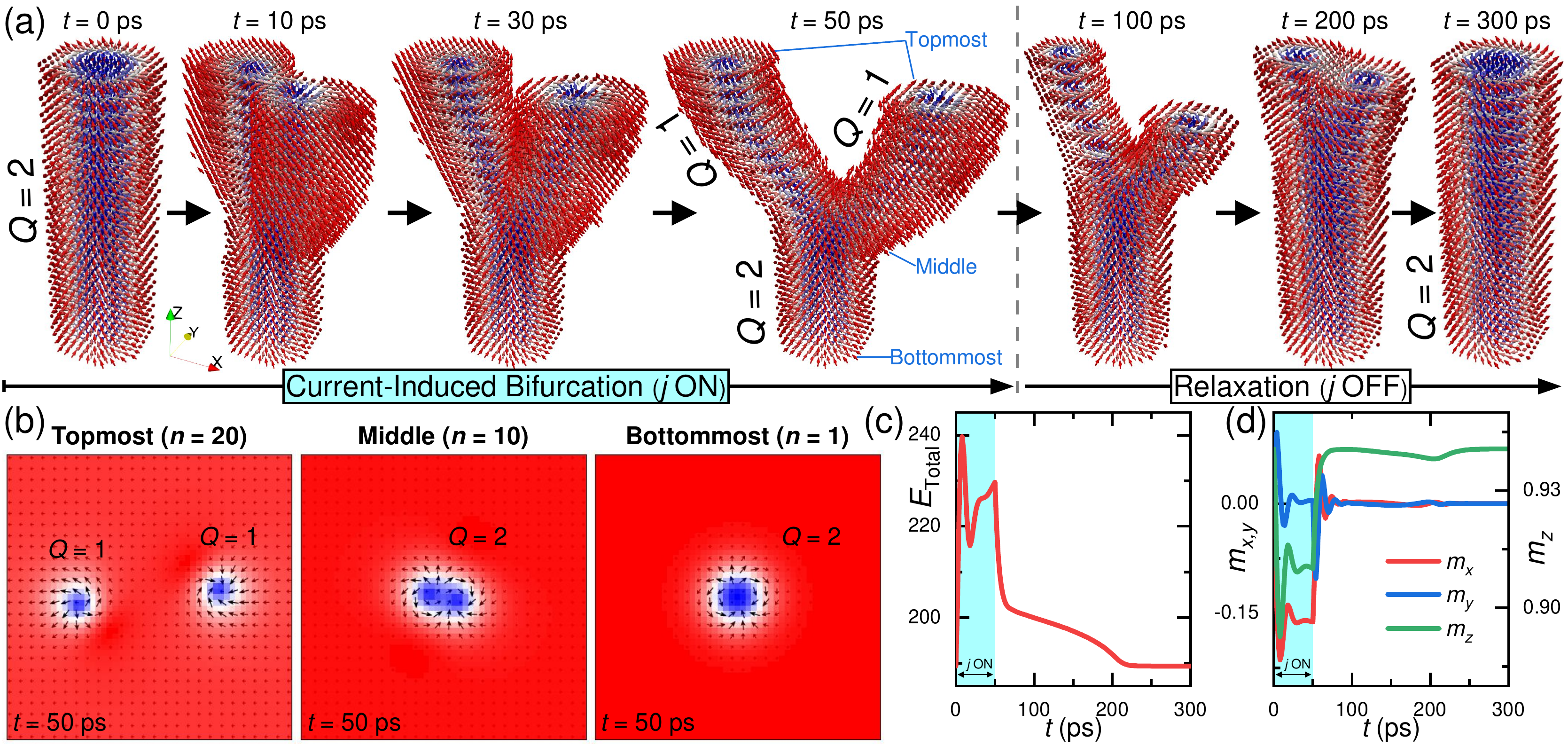}}
\caption{%
Bifurcation of an I-shaped skyrmion string with $Q=2$ to a Y-shaped skyrmion string with two $Q=1$ branches.
(a) Illustrations showing the bifurcation and relaxation at selected times. The driving current is turned on for $t=0-50$ ps.
(b) Top-view snapshots of the topmost, middle, and bottommost FM layers at $t=50$ ps. Black arrows represent spins with a subsample rate of $2$.
$n$ is the FM layer index~\cite{SI}.
(c) $E_{\text{Total}}$ as a function of time for the bifurcation and relaxation.
(d) Time-dependent $m_{x,y}$ and $m_z$ of the system for the bifurcation and relaxation.
}
\label{FIG2}
\end{figure*}
%%%%%%%%%%%%%%%%%%%%%%%%%%%%%%%%%%%%%%%%%%%%%%%%%%%%%%%%%%%%

%-%-%-%-%-%-%-%-%-%-%-%-%-%-%-%-%-%-%-%-%-%-%-%-%-%-%-%-%-%-%
%\section{Results and Discussion}
%\label{se:Results}
%-%-%-%-%-%-%-%-%-%-%-%-%-%-%-%-%-%-%-%-%-%-%-%-%-%-%-%-%-%-%

\textit{Separation of a 3D skyrmion string.}
We first show the possibility to separate an I-shaped skyrmion string into two isolated I-shaped skyrmion strings with $Q=1$ in the layered system.
As shown in Fig.~\ref{FIG1}(a), we place an I-shaped skyrmion string with $Q=2$ in the model, which is formed by $20$ aligned stacks of 2D skyrmions with $Q=2$ and same helicity [Fig.~\ref{FIG1}(b)] in the presence of interlayer exchange coupling between adjacent FM layers.
For each 2D skyrmion, it is parametrized as
$\boldsymbol{m}(\boldsymbol{r})=\boldsymbol{m}(\theta,\phi)=(\sin\theta\cos\phi,\sin\theta\sin\phi,\cos\theta)$,
where
$\phi=Q_{\text{v}}\varphi+\eta$
with $\varphi$ being the azimuthal angle ($0\le\varphi<2\pi$),
$Q_{\text{v}}=\frac{1}{2\pi}\oint_{C}d \phi$ being the vorticity,
and $\eta\in [0,2\pi)$ being the helicity.
$\theta$ rotates by an angle of $\pi$ for spins from the 2D skyrmion center to the 2D skyrmion edge in each FM layer~\cite{Zhang_JPCM2020,Gobel_PP2021}.
The I-shaped skyrmion string with $Q=2$ is relaxed as the initial state. It is a metastable state as each 2D skyrmion with $Q=2$ is a metastable state in the frustrated magnetic system~\cite{Xichao_NCOMMS2017,Xia_PRApplied2019,Zhang_PRB2020}. Note that the 2D skyrmion with $Q=2$ is usually unstable in magnets with asymmetric exchange interactions~\cite{Zhang_PRB2016}.

In frustrated magnetic systems, two 2D skyrmions with $Q=1$ may form a 2D skyrmion with $Q=2$ spontaneously as the energy of one skyrmion with $Q=2$ is smaller than the total energy of two skyrmions with $Q=1$~\cite{Xichao_NCOMMS2017,Xia_PRApplied2019,Zhang_PRB2020}. Such a feature implies that one may obtain two skyrmions with $Q=1$ from a skyrmion with $Q=2$ by external stimuli [Fig.~\ref{FIG1}(b)]~\cite{Xichao_NCOMMS2017,Xia_PRApplied2019}.
We find it is also possible to separate an I-shaped skyrmion string with $Q=2$ into two I-shaped skyrmion strings with $Q=1$ by applying the dampinglike spin-orbit torque in all FM layers.

To demonstrate the separation, we apply the spin current in all FM layers for $t=0-50$ ps with a current density $j=275$ MA cm$^{-2}$, and then relax the system at zero current for $350$ ps.
We find that the I-shaped skyrmion string with $Q=2$ is transformed into two I-shaped skyrmion strings due to the current-induced separation of 2D skyrmions in all FM layers [Fig.~\ref{FIG1}(e)] (see Supplemental Video \blue{1}~\cite{SI}).
The time-dependent system energy and spin components are given in Figs.~\ref{FIG1}(c) and~\ref{FIG1}(d), respectively, which suggest that the system with two skyrmion strings is relaxed soon after switching off the current. The spin components oscillate significantly during the separation and reach stable values through the relaxation. The system energy is higher than that of the initial state, which means the two skyrmion strings with $Q=1$ may merge spontaneously if they are close enough to each other.
In Fig.~\ref{FIG1}(f), we show the top-view snapshots of the topmost FM layer (i.e., $n=20$; $n$ is the FM layer index~\cite{SI}) at selected times. It shows that the two I-shaped skyrmion strings with $Q=1$ have opposite helicities, as illustrated schematically in Fig.~\ref{FIG1}(b). As the driving force and skyrmion structure are uniform in all FM layers, the time-dependent spin configurations are the same in the thickness dimension.

%%%%%%%%%%%%%%%%%%%%%%%%%%%%%%%%%%%%%%%%%%%%%%%%%%%%%%%%%%%%
\begin{figure*}[t]
\centerline{\includegraphics[width=1.00\textwidth]{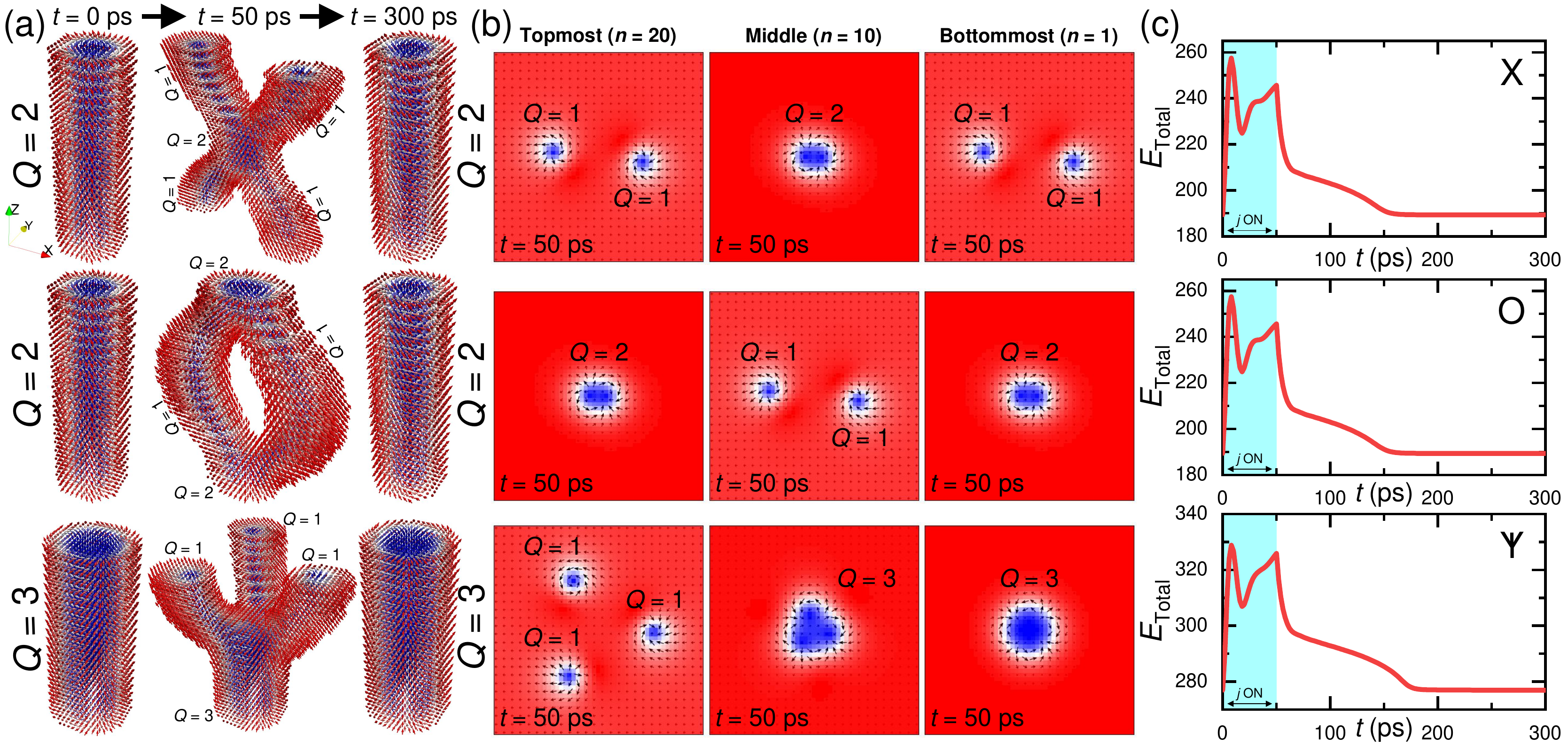}}
\caption{%
Bifurcation of an I-shaped skyrmion string with $Q=2$ or $3$ to an X-shaped, O-shaped, or Y-shaped skyrmion string.
(a) Illustrations showing the bifurcation and relaxation at selected times. The driving current is turned on for $t=0-50$ ps.
(b) Top-view snapshots of the topmost, middle, and bottommost FM layers at $t=50$ ps. Black arrows represent spins with a subsample rate of $2$.
(c) $E_{\text{Total}}$ as a function of time for the bifurcation and relaxation.
}
\label{FIG3}
\end{figure*}
%%%%%%%%%%%%%%%%%%%%%%%%%%%%%%%%%%%%%%%%%%%%%%%%%%%%%%%%%%%%

\textit{Bifurcation of a 3D skyrmion string.}
Based on the current-induced separation, we further demonstrate that it is possible to induce the bifurcation of the I-shaped skyrmion string by partially applying the dampinglike spin-orbit torque in the layered system.
In Fig.~\ref{FIG2}(a), the initial state at $t=0$ ps is a relaxed I-shaped skyrmion string with $Q=2$. We apply the spin current only in the top $10$ FM layers (i.e., $n=11-20$) with a current density $j=275$ MA cm$^{-2}$ for $t=0-50$ ps. The partially injected current leads to the separation of the skyrmion string with $Q=2$ into two skyrmion strings with $Q=1$ within the top $10$ FM layers, however, as the skyrmion string with $Q=2$ within the bottom $10$ FM layers is unchanged, a bifurcated Y-shaped skyrmion string is formed at $t=50$ ps.
Such a Y-shaped skyrmion string has two branches with $Q=1$ within the top half of the layered system and a trunk with $Q=2$ within the bottom half of the layered system.
Therefore, the region of connection of the trunk and two branches does not form a Bloch point~\cite{Malozemoff} to equalize the topological charges of the trunk ($Q=2$) and branches ($Q=1+1=2$); see Supplemental Fig.~\blue{3}~\cite{SI}.
Note that a Bloch point usually exists in the static Y-shaped skyrmion string in chiral bulk magnets~\cite{Seki_2021}, where both the trunk and two branches have $Q=1$ so that the connection of the trunk ($Q=1$) and branches ($Q=1+1=2$) must involve a Bloch point with $Q=-1$.
When the current is turned off, the two branches with $Q=1$ are merged into one skyrmion string with $Q=2$ during the relaxation. As a result, the bifurcation is recovered and the Y-shaped skyrmion string is relaxed to an I-shaped skyrmion string before $t=300$ ps.

In Fig.~\ref{FIG2}(b), we show the top-view snapshots of the topmost, middle, and bottommost FM layers at $t=50$ ps, where the Y-shaped skyrmion string is most obvious. It shows that the two skyrmion branches have clockwise and counterclockwise helicities, respectively.
The in-layer skyrmion structure at the junction of the Y-shaped skyrmion string (i.e., $n=10$) show a peanut-shaped skyrmion structure with $Q=2$, which may be regarded as a biskyrmion structure~\cite{Yu_NC2014,Gobel_SR2019,Capic_PRR2019}.
The time-dependent system energy and spin components are given in Figs.~\ref{FIG2}(c) and~\ref{FIG2}(d), respectively.
The merging of the two bifurcated skyrmion branches during the relaxation is accompanied by the rotation of the two branches with respect to the trunk axis; see Supplemental Video \blue{2}~\cite{SI}.
From the time-dependent energy curve, it can be seen that the relaxation after $t=50$ ps (i.e., the spontaneous recovery of bifurcation) is slower than that of two separated I-shaped skyrmion strings with $Q=1$ due to the rotation [cf. Fig.~\ref{FIG1}(c)].
A reversed Y-shaped skyrmion string could be created in a similar way (see Supplemental Fig.~\blue{4} and Supplemental Video \blue{3}~\cite{SI}).

In Fig.~\ref{FIG3}(a), we demonstrate the creation of an X-shaped skyrmion string with four $Q=1$ branches and an O-shaped skyrmion string with two $Q=1$ branches by inducing the bifurcation of an I-shaped skyrmion string with $Q=2$ via partially injected spin current.
We also create a Y-shaped skyrmion string with three $Q=1$ branches from an I-shaped skyrmion string with $Q=3$.
To create an X-shaped skyrmion string (see Supplemental Video \blue{4}~\cite{SI}), the current is only injected into the top $7$ FM layers (i.e., $n=14-20$) and bottom $7$ FM layers (i.e., $n=1-7$) for $t=0-50$ ps.
To create an O-shaped skyrmion string (see Supplemental Video \blue{5}~\cite{SI}), the current is only injected into the middle $14$ FM layers (i.e., $n=4-17$) for $t=0-50$ ps.
To create a Y-shaped skyrmion string with three $Q=1$ branches (see Supplemental Video \blue{6}~\cite{SI}), the current is injected into the top $10$ FM layers (i.e., $n=11-20$) for $t=0-50$ ps, where a relaxed metastable I-shaped skyrmion string with $Q=3$ is the initial state.

In Fig.~\ref{FIG3}(b), we show the top-view snapshots of the topmost, middle, and bottommost FM layers at $t=50$ ps, where the bifurcated skyrmion strings are most obvious.
Similar to the bifurcation of an I-shaped skyrmion string to a Y-shaped skyrmion string [Fig.~\ref{FIG2}], the bifurcated skyrmion branches have different helicities in the same FM layer.
The time-dependent total energies for the bifurcation and relaxation corresponding to Fig.~\ref{FIG3}(a) are given in Fig.~\ref{FIG3}(c). The formation and recovery of X-shaped and O-shaped skyrmion strings show almost identical energy curves. The reason is that $14$ FM layers in total are driven by the same spin current for both the two cases.

%-%-%-%-%-%-%-%-%-%-%-%-%-%-%-%-%-%-%-%-%-%-%-%-%-%-%-%-%-%-%
%\section{Conclusion}
%\label{se:Conclusion}
%-%-%-%-%-%-%-%-%-%-%-%-%-%-%-%-%-%-%-%-%-%-%-%-%-%-%-%-%-%-%

\textit{Conclusion and outlook.}
In conclusion, we have studied the current-induced separation and bifurcation of a 3D skyrmion string carrying a topological charge of $Q=2$ or $3$.
We find that the bifurcation could lead to the formation of Y-shaped, X-shaped, and O-shaped skyrmion strings, which depends on the current injection geometry in the layered frustrated magnetic system.
The Y-shaped, X-shaped, and O-shaped skyrmion strings have multiple skyrmion branches with $Q=1$ and opposite helicities.
The bifurcated skyrmion string could spontaneously recover an I-shaped skyrmion string when the driving current is turned off, where the branches show rotation with respect to the trunk axis.

It is noteworthy that the speed of the branch rotation as well as the overall recovery speed can be reduced by applying a small current during the relaxation; see Supplemental Fig.~\blue{5}~\cite{SI}. Such a feature means that the lifetime of a bifurcated skyrmion string can be controlled electrically. This also means one can switch on and off the bifurcated state and manipulate it by electrical means.
Hence, it is possible to design a multistate memory unit based on a skyrmion string in a three-terminal magnetic tunnel junction, where the information is encoded by the number of skyrmion branches and is electrically controlled by current pulses.
Indeed, the current-induced skyrmion string bifurcation may also be used for controlling the propagation of magnons along skyrmion strings in a 3D spintronic building block.

On the other hand, for the bifurcation of a skyrmion string with $Q=2$, we note that the delocalization of the driving force in the thickness dimension may result in the incomplete bifurcation or the fracture of bifurcated branches; see Supplemental Fig.~\blue{6}~\cite{SI}.
Besides, the reversed current propagation direction would not result in a qualitative change of the bifurcated skyrmion string; see Supplemental Fig.~\blue{7}~\cite{SI}.
The mechanism of bifurcation should be universal to any stable or metastable skyrmion strings with $|Q|\geq 2$.
Our results reveal unusual dynamic physics of 3D skyrmion strings carrying high topological charges.

%-%-%-%-%-%-%-%-%-%-%-%-%-%-%-%-%-%-%-%-%-%-%-%-%-%-%-%-%-%-%
\begin{acknowledgments}
%
%\textit{Acknowledgments.}
%
J.X. acknowledges support by the National Natural Science Foundation of China (Grant No. 12104327).
X.Z. was an International Research Fellow of the Japan Society for the Promotion of Science (JSPS). X.Z. was supported by JSPS KAKENHI (Grant No. JP20F20363).
O.A.T. acknowledges support from the Australian Research Council (Grant No. DP200101027), Russian Science Foundation (Project No. 21-79-20186), NCMAS grant, and the Cooperative Research Project Program at the Research Institute of Electrical Communication, Tohoku University.
J.Y. acknowledges support by the National Natural Science Foundation of China (Grant No. 51731001).
G.Z. acknowledges support by the National Natural Science Foundation of China (Grant Nos. 51771127, 52171188, and 52111530143).
M.E. acknowledges support by the Grants-in-Aid for Scientific Research from JSPS KAKENHI (Grant Nos. JP18H03676 and JP17K05490) and the support by CREST, JST (Grant Nos. JPMJCR20T2 and JPMJCR16F1).
Y.Z. acknowledges support by the Guangdong Special Support Project (Grant No. 2019BT02X030), Shenzhen Fundamental Research Fund (Grant No. JCYJ20210324120213037), Shenzhen Peacock Group Plan (Grant No. KQTD20180413181702403), Pearl River Recruitment Program of Talents (Grant No. 2017GC010293), and National Natural Science Foundation of China (Grant Nos. 11974298, 12004320, and 61961136006).
X.L. acknowledges support by the Grants-in-Aid for Scientific Research from JSPS KAKENHI (Grant Nos. JP20F20363, JP21H01364, and JP21K18872).
\end{acknowledgments}
%-%-%-%-%-%-%-%-%-%-%-%-%-%-%-%-%-%-%-%-%-%-%-%-%-%-%-%-%-%-%

%-%-%-%-%-%-%-%-%-%-%-%-%-%-%-%-%-%-%-%-%-%-%-%-%-%-%-%-%-%-%

%-%-%-%-%-%-%-%-%-%-%-%-%-%-%-%-%-%-%-%-%-%-%-%-%-%-%-%-%-%-%

%%%%%%%%%%%%%%%%%%%%%%%%%%%%%%%%%%%%%%%%%%%%%%%%%%%%%%%%%%%%
\end{document}